\documentclass[%
 preprint,
 showpacs,
 showkeys,
 preprintnumbers,
 amsmath,amssymb,
 aps,
  prl,
  longbibliography,
 ]{revtex4-1}

\usepackage[breaklinks=true,colorlinks=true,anchorcolor=blue,citecolor=blue,filecolor=blue,menucolor=blue,pagecolor=blue,urlcolor=blue,linkcolor=blue]{hyperref}
\usepackage{graphicx}
\usepackage{xcolor}
\usepackage[left]{eurosym}
\usepackage{url}

\begin{document}

\title{Neutrino dispersion relation changes due to radiative corrections as the origin of faster-than-light-in-vacuum propagation in a medium}


\author{Karl Svozil}
\affiliation{Institute of Theoretical Physics, Vienna
    University of Technology, Wiedner Hauptstra\ss e 8-10/136, A-1040
    Vienna, Austria}
\email{svozil@tuwien.ac.at} \homepage[]{http://tph.tuwien.ac.at/~svozil}

\preprint{CDMTCS preprint nr. 407/2011}

\begin{abstract}
Radiative corrections to the dispersion of neutrinos in nonstandard vacuum may give rise to ``boosts'' in their speed. This could explain recent experimental evidence by the OPERA collaboration, as well as the null result indicated by the supernova 1987A (SN 1987A) measurements of neutrino and photon arrival times.
\end{abstract}

\maketitle

The speed of light in a medium~\cite{svozil-putz-sol}
or under certain geometric constraints~\cite{scharnhorst,milonni,Scharnhorst-1998,barton,0305-4470-26-8-024}
may be decreased by radiative corrections with respect to vacuum values.
``Diagrammatically speaking''~\cite{feynman-62,schweber-62,thooft-Veltman_Diagrammar}, i.e., in terms of perturbative quantum field theory,
a quantum travels through the vacuum ether medium~\cite{dirac-aether}
through partly ``splitting up'' into,  and recombining from virtual particles.
Thus, intuitively speaking, it may not be too unreasonable to conceive such a particle  spending ``some of its existence''
in these virtual particle states.
As many of these virtual particles, although off shell,
are massive, this may temporarily decelerate a massless, or almost massless  quantum like a neutrino
as it traverses a nonvacuum medium such as the matter crossed in the OPERA experiment~\cite{Opera-Cern}.

At the same time, this may explain why for neutrinos from a cosmic source,
the coincidence in time with the optical sighting of the source
has been consistent
with a neutrino speed  comparable to the speed of light in vacuum \cite{PhysRevD.38.448}.
Because in the case of a distant supernova event,
the vacuum medium traversed by both types of quanta has been identical
for both
particle types -- the cosmic neutrinos and photons registered on Earth -- respectively.

Thus, any change of vacuum polarization, such as finite boundary conditions, or increased or decreased pair production,
alters the susceptibility of the vacuum ether medium for carrying quanta,
and thus results in a change of the velocity of these particles.
Analogue situations have been investigated
for magnetic fields~\cite{er:61,RevModPhys.38.626,Adler1971599} and finite temperatures~\cite{Gies1998420}.
A vacuum polarization-induced index of refraction {\em smaller than one} was reported by
Scharnhorst~\cite{scharnhorst,milonni,Scharnhorst-1998} and Barton~\cite{barton,0305-4470-26-8-024}
in an attempt to utilize the reduced vacuum polarization
in the  Casimir vacuum~\cite{milonni-book} between two conducting parallel plates.
Moreover, trans-vacuum-speed
metamaterials~\cite{PhysRevE.63.046604,PhysRevE.68.026612,PhysRevE.70.068601,PhysRevE.70.068602,PhysRevE.78.016601}
as well as negative refractive indices in gyrotropically magnetoelectric media~\cite{PhysRevB.75.196101} have been suggested.

One of the possibilities which have  been discussed recently~\cite{svozil-putz-sol} is the immersion of photons
into a vacuum ether medium ``occupied'' by electrons or positrons~\cite{svozil-putz-sol}.
In such an environment, the Pauli exclusion principle would ``attenuate'' pair creation, thereby reducing the  polarization of the medium, resulting in
a reduced index of refraction as well as in an increase of the velocity of light.

The lowest order  change
to the radiative correction associated with  the vacuum polarization of the photon
can be written as~\cite{RevModPhys.21.434,PhysRev.76.769,schweber-62}
\begin{equation}
{\rm \Delta }{\bf \Pi}_{\mu \nu}(k^2)=-\left(g_{\mu \nu }k^2 - k_\mu k_\nu \right) \frac{2\alpha}{3\pi}  \log \frac{\varepsilon_F}{m}.
\end{equation}
Here $m$ stands for the electron rest mass and $\varepsilon_F$ denotes the cutoff
associated with the filled electron or positron modes.
Let $\epsilon_\mu$ stand for the vacuum polarization.
Then an effective mass term can be introduced~\cite{PhysRev.82.664,PhysRevD.10.492,PhysRevD.12.1132}
\begin{equation}
M(k)=\epsilon^\mu {\bf \Pi}_{\mu \nu}(k)\epsilon^\nu
\end{equation}
such that the eigenvalue equation is
\begin{equation}
{\bf k}^2+ M(k)=(k^0)^2,
\end{equation}
where $k^\mu=({\bf k},k^0=\omega)$; and
\begin{equation}
\vert  {\bf k} \vert \approx \omega - \frac{1}{2 \omega} M(k).
\end{equation}
The index of refraction may be defined by
\begin{equation}
n(\omega )=\frac{\vert {\bf k} \vert}{\omega}\approx 1 - \frac{1}{2 \omega^2} M(k).
\end{equation}
Hence the change of the refractive index is given by
\begin{equation}
{\rm \Delta }n(\omega )\approx -\frac{\alpha}{3\pi \omega^2} (\epsilon^\mu k_\mu)^2  \log \frac{\varepsilon_F}{m}.
\label{2010-uc10-e1}
\end{equation}

The group velocity is given by~\cite[Equ.~(2)]{Scharnhorst-1998} $v_{gr}=c/n_{gr}$ with
$n_{gr}(\omega )= n (\omega )+ \omega \left[\partial n (\omega )/\partial \omega \right]$,
which, for transversal waves, turns out to be $n (\omega )$.
As a result, the speed of photons in such a medium $c/(1-{\rm \Delta }n)\approx c+{\rm \Delta }c$  exceeds the velocity of light in vacuum by
${\rm \Delta }c = c {\rm \Delta }n$.
Thereby it should always be kept in mind that group velocities,
like phase velocities and energy velocities, in general are not signal velocities; hence
group velocities exceeding the vacuum speed of light $c$ does
not contradict relativity
\cite{PhysRevA.48.R34,Diener1996327,Chiao:02}.

In summary we have discussed field theoretic options for the ``speedup'' of ultrarelativistic particles beyond the speed of light barrier in
the presence of suitable media which cause a reduction
of polarizability and radiative corrections.
These considerations do neither represent the possibility to circumvent relativistic causality,
because no {\em ad hoc} ``willable'' superluminal information or paradoxical time travel will be rendered~\cite{svozil-greenberger-2005};
nor are they inconsistent with the present formalism of relativity theory or the theory of quantized fields;
on the contrary they can be taken as a demonstration of the relativistic formalism~\cite{svozil-relrel,svozil-2001-convention}.
They would not even make necessary the standard SI conventionalization of the constancy of the speed of light~\cite{peres-84};
with the possible addendum of referring to this the velocity of light in a particular type of rather idealized vacuum.


\begin{thebibliography}{35}%
\makeatletter
\providecommand \@ifxundefined [1]{%
 \@ifx{#1\undefined}
}%
\providecommand \@ifnum [1]{%
 \ifnum #1\expandafter \@firstoftwo
 \else \expandafter \@secondoftwo
 \fi
}%
\providecommand \@ifx [1]{%
 \ifx #1\expandafter \@firstoftwo
 \else \expandafter \@secondoftwo
 \fi
}%
\providecommand \natexlab [1]{#1}%
\providecommand \enquote  [1]{``#1''}%
\providecommand \bibnamefont  [1]{#1}%
\providecommand \bibfnamefont [1]{#1}%
\providecommand \citenamefont [1]{#1}%
\providecommand \href@noop [0]{\@secondoftwo}%
\providecommand \href [0]{\begingroup \@sanitize@url \@href}%
\providecommand \@href[1]{\@@startlink{#1}\@@href}%
\providecommand \@@href[1]{\endgroup#1\@@endlink}%
\providecommand \@sanitize@url [0]{\catcode `\\12\catcode `\$12\catcode
  `\&12\catcode `\#12\catcode `\^12\catcode `\_12\catcode `\%12\relax}%
\providecommand \@@startlink[1]{}%
\providecommand \@@endlink[0]{}%
\providecommand \url  [0]{\begingroup\@sanitize@url \@url }%
\providecommand \@url [1]{\endgroup\@href {#1}{\urlprefix }}%
\providecommand \urlprefix  [0]{URL }%
\providecommand \Eprint [0]{\href }%
\providecommand \doibase [0]{http://dx.doi.org/}%
\providecommand \selectlanguage [0]{\@gobble}%
\providecommand \bibinfo  [0]{\@secondoftwo}%
\providecommand \bibfield  [0]{\@secondoftwo}%
\providecommand \translation [1]{[#1]}%
\providecommand \BibitemOpen [0]{}%
\providecommand \bibitemStop [0]{}%
\providecommand \bibitemNoStop [0]{.\EOS\space}%
\providecommand \EOS [0]{\spacefactor3000\relax}%
\providecommand \BibitemShut  [1]{\csname bibitem#1\endcsname}%
\let\auto@bib@innerbib\@empty
\bibitem [{\citenamefont {Putz}\ and\ \citenamefont
  {Svozil}(2010)}]{svozil-putz-sol}%
  \BibitemOpen
  \bibfield  {author} {\bibinfo {author} {\bibfnamefont {Volkmar}\ \bibnamefont
  {Putz}}\ and\ \bibinfo {author} {\bibfnamefont {Karl}\ \bibnamefont
  {Svozil}},\ }\bibfield  {title} {\enquote {\bibinfo {title} {Can a computer
  be ``pushed'' to perform faster-than-light?}}\ }\href
  {http://arxiv.org/abs/1003.1238} {\  (\bibinfo {year} {2010})},\ \Eprint
  {http://arxiv.org/abs/arXiv:1003.1238} {arXiv:1003.1238} \BibitemShut
  {NoStop}%
\bibitem [{\citenamefont {Scharnhorst}(1990)}]{scharnhorst}%
  \BibitemOpen
  \bibfield  {author} {\bibinfo {author} {\bibfnamefont {K.}~\bibnamefont
  {Scharnhorst}},\ }\bibfield  {title} {\enquote {\bibinfo {title} {On
  propagation of light in the vacuum between plates},}\ }\href {\doibase
  10.1016/0370-2693(90)90997-K} {\bibfield  {journal} {\bibinfo  {journal}
  {Physics Letters B}\ }\textbf {\bibinfo {volume} {236}},\ \bibinfo {pages}
  {354--359} (\bibinfo {year} {1990})}\BibitemShut {NoStop}%
\bibitem [{\citenamefont {Milonni}\ and\ \citenamefont
  {Svozil}(1990)}]{milonni}%
  \BibitemOpen
  \bibfield  {author} {\bibinfo {author} {\bibfnamefont {Peter}\ \bibnamefont
  {Milonni}}\ and\ \bibinfo {author} {\bibfnamefont {Karl}\ \bibnamefont
  {Svozil}},\ }\bibfield  {title} {\enquote {\bibinfo {title} {Impossibility of
  measuring faster-than-c signaling by the {S}charnhorst effect},}\ }\href
  {\doibase 10.1016/0370-2693(90)90317-Y} {\bibfield  {journal} {\bibinfo
  {journal} {Physics Letters B}\ }\textbf {\bibinfo {volume} {248}},\ \bibinfo
  {pages} {437--438} (\bibinfo {year} {1990})}\BibitemShut {NoStop}%
\bibitem [{\citenamefont {Scharnhorst}(1998)}]{Scharnhorst-1998}%
  \BibitemOpen
  \bibfield  {author} {\bibinfo {author} {\bibfnamefont {K.}~\bibnamefont
  {Scharnhorst}},\ }\bibfield  {title} {\enquote {\bibinfo {title} {The
  velocities of light in modified {QED} vacua},}\ }\href {\doibase
  10.1002/(SICI)1521-3889(199812)7:7/8<700::AID-ANDP700>3.0.CO;2-K} {\bibfield
  {journal} {\bibinfo  {journal} {Annalen der Physik}\ }\textbf {\bibinfo
  {volume} {7}},\ \bibinfo {pages} {700--709} (\bibinfo {year} {1998})},\
  \Eprint {http://arxiv.org/abs/arXiv:hep-th/9810221} {arXiv:hep-th/9810221}
  \BibitemShut {NoStop}%
\bibitem [{\citenamefont {Barton}(1990)}]{barton}%
  \BibitemOpen
  \bibfield  {author} {\bibinfo {author} {\bibfnamefont {G.}~\bibnamefont
  {Barton}},\ }\bibfield  {title} {\enquote {\bibinfo {title} {Faster-than-c
  light between parallel mirrors. {T}he {S}charnhorst effect rederived},}\
  }\href {\doibase 10.1016/0370-2693(90)91224-Y} {\bibfield  {journal}
  {\bibinfo  {journal} {Physics Letters B}\ }\textbf {\bibinfo {volume}
  {237}},\ \bibinfo {pages} {559--562} (\bibinfo {year} {1990})}\BibitemShut
  {NoStop}%
\bibitem [{\citenamefont {Barton}\ and\ \citenamefont
  {Scharnhorst}(1993)}]{0305-4470-26-8-024}%
  \BibitemOpen
  \bibfield  {author} {\bibinfo {author} {\bibfnamefont {G.}~\bibnamefont
  {Barton}}\ and\ \bibinfo {author} {\bibfnamefont {K.}~\bibnamefont
  {Scharnhorst}},\ }\bibfield  {title} {\enquote {\bibinfo {title} {{QED}
  between parallel mirrors: light signals faster than c, or amplified by the
  vacuum},}\ }\href {\doibase 10.1088/0305-4470/26/8/024} {\bibfield  {journal}
  {\bibinfo  {journal} {Journal of Physics A: Mathematical and General}\
  }\textbf {\bibinfo {volume} {26}},\ \bibinfo {pages} {2037--2046} (\bibinfo
  {year} {1993})}\BibitemShut {NoStop}%
\bibitem [{\citenamefont {Feynman}(1962)}]{feynman-62}%
  \BibitemOpen
  \bibfield  {author} {\bibinfo {author} {\bibfnamefont {Richard~Phillips}\
  \bibnamefont {Feynman}},\ }\href@noop {} {\emph {\bibinfo {title} {Quantum
  Electrodynamics}}}\ (\bibinfo  {publisher} {Addison-Wesley},\ \bibinfo
  {address} {Redwood City, CA},\ \bibinfo {year} {1962})\BibitemShut {NoStop}%
\bibitem [{\citenamefont {Schweber}(1984)}]{schweber-62}%
  \BibitemOpen
  \bibfield  {author} {\bibinfo {author} {\bibfnamefont {Silvan}\ \bibnamefont
  {Schweber}},\ }\href@noop {} {\emph {\bibinfo {title} {Relativistic Quantum
  Field Theory}}}\ (\bibinfo  {publisher} {Harper and Row},\ \bibinfo {address}
  {New York},\ \bibinfo {year} {1984})\BibitemShut {NoStop}%
\bibitem [{\citenamefont {`t~Hooft}\ and\ \citenamefont
  {Veltman}(1973)}]{thooft-Veltman_Diagrammar}%
  \BibitemOpen
  \bibfield  {author} {\bibinfo {author} {\bibfnamefont {G.}~\bibnamefont
  {`t~Hooft}}\ and\ \bibinfo {author} {\bibfnamefont {M.}~\bibnamefont
  {Veltman}},\ }\href {http://doc.cern.ch/yellowrep/1973/1973-009/p1.pdf}
  {\enquote {\bibinfo {title} {Diagrammar},}\ } (\bibinfo {year} {1973}),\
  \bibinfo {note} {cERN preprint 73-9}\BibitemShut {NoStop}%
\bibitem [{\citenamefont {Dirac}(1951)}]{dirac-aether}%
  \BibitemOpen
  \bibfield  {author} {\bibinfo {author} {\bibfnamefont {Paul Adrien~Maurice}\
  \bibnamefont {Dirac}},\ }\bibfield  {title} {\enquote {\bibinfo {title} {Is
  there an aether?}}\ }\href {\doibase 10.1038/168906a0} {\bibfield  {journal}
  {\bibinfo  {journal} {Nature}\ }\textbf {\bibinfo {volume} {168}},\ \bibinfo
  {pages} {906--907} (\bibinfo {year} {1951})}\BibitemShut {NoStop}%
\bibitem [{\citenamefont {Collaboraton}(2011)}]{Opera-Cern}%
  \BibitemOpen
  \bibfield  {author} {\bibinfo {author} {\bibfnamefont {The~OPERA}\
  \bibnamefont {Collaboraton}},\ }\href@noop {} {\enquote {\bibinfo {title}
  {Measurement of the neutrino velocity with the opera detector in the {CNGS}
  beam},}\ } (\bibinfo {year} {2011}),\ \Eprint
  {http://arxiv.org/abs/arXiv:1109.4897} {arXiv:1109.4897} \BibitemShut
  {NoStop}%
\bibitem [{\citenamefont {Hirata}\ \emph {et~al.}(1988)\citenamefont {Hirata},
  \citenamefont {Kajita}, \citenamefont {Koshiba}, \citenamefont {Nakahata},
  \citenamefont {Oyama}, \citenamefont {Sato}, \citenamefont {Suzuki},
  \citenamefont {Takita}, \citenamefont {Totsuka}, \citenamefont {Kifune},
  \citenamefont {Suda}, \citenamefont {Takahashi}, \citenamefont {Tanimori},
  \citenamefont {Miyano}, \citenamefont {Yamada}, \citenamefont {Beier},
  \citenamefont {Feldscher}, \citenamefont {Frati}, \citenamefont {Kim},
  \citenamefont {Mann}, \citenamefont {Newcomer}, \citenamefont {Van~Berg},
  \citenamefont {Zhang},\ and\ \citenamefont {Cortez}}]{PhysRevD.38.448}%
  \BibitemOpen
  \bibfield  {author} {\bibinfo {author} {\bibfnamefont {K.~S.}\ \bibnamefont
  {Hirata}}, \bibinfo {author} {\bibfnamefont {T.}~\bibnamefont {Kajita}},
  \bibinfo {author} {\bibfnamefont {M.}~\bibnamefont {Koshiba}}, \bibinfo
  {author} {\bibfnamefont {M.}~\bibnamefont {Nakahata}}, \bibinfo {author}
  {\bibfnamefont {Y.}~\bibnamefont {Oyama}}, \bibinfo {author} {\bibfnamefont
  {N.}~\bibnamefont {Sato}}, \bibinfo {author} {\bibfnamefont {A.}~\bibnamefont
  {Suzuki}}, \bibinfo {author} {\bibfnamefont {M.}~\bibnamefont {Takita}},
  \bibinfo {author} {\bibfnamefont {Y.}~\bibnamefont {Totsuka}}, \bibinfo
  {author} {\bibfnamefont {T.}~\bibnamefont {Kifune}}, \bibinfo {author}
  {\bibfnamefont {T.}~\bibnamefont {Suda}}, \bibinfo {author} {\bibfnamefont
  {K.}~\bibnamefont {Takahashi}}, \bibinfo {author} {\bibfnamefont
  {T.}~\bibnamefont {Tanimori}}, \bibinfo {author} {\bibfnamefont
  {K.}~\bibnamefont {Miyano}}, \bibinfo {author} {\bibfnamefont
  {M.}~\bibnamefont {Yamada}}, \bibinfo {author} {\bibfnamefont {E.~W.}\
  \bibnamefont {Beier}}, \bibinfo {author} {\bibfnamefont {L.~R.}\ \bibnamefont
  {Feldscher}}, \bibinfo {author} {\bibfnamefont {W.}~\bibnamefont {Frati}},
  \bibinfo {author} {\bibfnamefont {S.~B.}\ \bibnamefont {Kim}}, \bibinfo
  {author} {\bibfnamefont {A.~K.}\ \bibnamefont {Mann}}, \bibinfo {author}
  {\bibfnamefont {F.~M.}\ \bibnamefont {Newcomer}}, \bibinfo {author}
  {\bibfnamefont {R.}~\bibnamefont {Van~Berg}}, \bibinfo {author}
  {\bibfnamefont {W.}~\bibnamefont {Zhang}}, \ and\ \bibinfo {author}
  {\bibfnamefont {B.~G.}\ \bibnamefont {Cortez}},\ }\bibfield  {title}
  {\enquote {\bibinfo {title} {Observation in the {K}amiokande-{II} detector of
  the neutrino burst from supernova {SN1987A}},}\ }\href {\doibase
  10.1103/PhysRevD.38.448} {\bibfield  {journal} {\bibinfo  {journal} {Physical
  Review D}\ }\textbf {\bibinfo {volume} {38}},\ \bibinfo {pages} {448--458}
  (\bibinfo {year} {1988})}\BibitemShut {NoStop}%
\bibitem [{\citenamefont {Erber}(1961)}]{er:61}%
  \BibitemOpen
  \bibfield  {author} {\bibinfo {author} {\bibfnamefont {Thomas}\ \bibnamefont
  {Erber}},\ }\bibfield  {title} {\enquote {\bibinfo {title} {Velocity of light
  in a magnetic field},}\ }\href {\doibase 10.1038/190025a0} {\bibfield
  {journal} {\bibinfo  {journal} {Nature}\ }\textbf {\bibinfo {volume} {190}},\
  \bibinfo {pages} {25--27} (\bibinfo {year} {1961})}\BibitemShut {NoStop}%
\bibitem [{\citenamefont {Erber}(1966)}]{RevModPhys.38.626}%
  \BibitemOpen
  \bibfield  {author} {\bibinfo {author} {\bibfnamefont {Thomas}\ \bibnamefont
  {Erber}},\ }\bibfield  {title} {\enquote {\bibinfo {title} {High-energy
  electromagnetic conversion processes in intense magnetic fields},}\ }\href
  {\doibase 10.1103/RevModPhys.38.626} {\bibfield  {journal} {\bibinfo
  {journal} {Reviews of Modern Physics}\ }\textbf {\bibinfo {volume} {38}},\
  \bibinfo {pages} {626--659} (\bibinfo {year} {1966})}\BibitemShut {NoStop}%
\bibitem [{\citenamefont {Adler}(1971)}]{Adler1971599}%
  \BibitemOpen
  \bibfield  {author} {\bibinfo {author} {\bibfnamefont {Stephen~L.}\
  \bibnamefont {Adler}},\ }\bibfield  {title} {\enquote {\bibinfo {title}
  {Photon splitting and photon dispersion in a strong magnetic field},}\ }\href
  {\doibase 10.1016/0003-4916(71)90154-0} {\bibfield  {journal} {\bibinfo
  {journal} {Annals of Physics}\ }\textbf {\bibinfo {volume} {67}},\ \bibinfo
  {pages} {599--647} (\bibinfo {year} {1971})}\BibitemShut {NoStop}%
\bibitem [{\citenamefont {Gies}\ and\ \citenamefont
  {Dittrich}(1998)}]{Gies1998420}%
  \BibitemOpen
  \bibfield  {author} {\bibinfo {author} {\bibfnamefont {Holger}\ \bibnamefont
  {Gies}}\ and\ \bibinfo {author} {\bibfnamefont {Walter}\ \bibnamefont
  {Dittrich}},\ }\bibfield  {title} {\enquote {\bibinfo {title} {Light
  propagation in non-trivial {QED} vacua},}\ }\href {\doibase
  10.1016/S0370-2693(98)00572-3} {\bibfield  {journal} {\bibinfo  {journal}
  {Physics Letters B}\ }\textbf {\bibinfo {volume} {431}},\ \bibinfo {pages}
  {420 -- 429} (\bibinfo {year} {1998})}\BibitemShut {NoStop}%
\bibitem [{\citenamefont {Milonni}(1994)}]{milonni-book}%
  \BibitemOpen
  \bibfield  {author} {\bibinfo {author} {\bibfnamefont {Peter~W.}\
  \bibnamefont {Milonni}},\ }\href@noop {} {\emph {\bibinfo {title} {The
  Quantum Vacuum}}}\ (\bibinfo  {publisher} {Academic Press},\ \bibinfo
  {address} {San Diego},\ \bibinfo {year} {1994})\BibitemShut {NoStop}%
\bibitem [{\citenamefont {Ziolkowski}(2001)}]{PhysRevE.63.046604}%
  \BibitemOpen
  \bibfield  {author} {\bibinfo {author} {\bibfnamefont {Richard~W.}\
  \bibnamefont {Ziolkowski}},\ }\bibfield  {title} {\enquote {\bibinfo {title}
  {Superluminal transmission of information through an electromagnetic
  metamaterial},}\ }\href {\doibase 10.1103/PhysRevE.63.046604} {\bibfield
  {journal} {\bibinfo  {journal} {Physical Review E}\ }\textbf {\bibinfo
  {volume} {63}},\ \bibinfo {pages} {046604} (\bibinfo {year}
  {2001})}\BibitemShut {NoStop}%
\bibitem [{\citenamefont {Ziolkowski}\ and\ \citenamefont
  {Cheng}(2003)}]{PhysRevE.68.026612}%
  \BibitemOpen
  \bibfield  {author} {\bibinfo {author} {\bibfnamefont {Richard~W.}\
  \bibnamefont {Ziolkowski}}\ and\ \bibinfo {author} {\bibfnamefont
  {Ching-Ying}\ \bibnamefont {Cheng}},\ }\bibfield  {title} {\enquote {\bibinfo
  {title} {Existence and design of trans-vacuum-speed metamaterials},}\ }\href
  {\doibase 10.1103/PhysRevE.68.026612} {\bibfield  {journal} {\bibinfo
  {journal} {Physical Review E}\ }\textbf {\bibinfo {volume} {68}},\ \bibinfo
  {pages} {026612} (\bibinfo {year} {2003})}\BibitemShut {NoStop}%
\bibitem [{\citenamefont {Tretyakov}(2004)}]{PhysRevE.70.068601}%
  \BibitemOpen
  \bibfield  {author} {\bibinfo {author} {\bibfnamefont {S.~A.}\ \bibnamefont
  {Tretyakov}},\ }\bibfield  {title} {\enquote {\bibinfo {title} {Comment on
  ``existence and design of trans-vacuum-speed metamaterials''},}\ }\href
  {\doibase 10.1103/PhysRevE.70.068601} {\bibfield  {journal} {\bibinfo
  {journal} {Physical Review E}\ }\textbf {\bibinfo {volume} {70}},\ \bibinfo
  {pages} {068601} (\bibinfo {year} {2004})}\BibitemShut {NoStop}%
\bibitem [{\citenamefont {Ziolkowski}(2004)}]{PhysRevE.70.068602}%
  \BibitemOpen
  \bibfield  {author} {\bibinfo {author} {\bibfnamefont {Richard~W.}\
  \bibnamefont {Ziolkowski}},\ }\bibfield  {title} {\enquote {\bibinfo {title}
  {Reply to ``comment on `existence and design of trans-vacuum-speed
  metamaterials' ''},}\ }\href {\doibase 10.1103/PhysRevE.70.068602} {\bibfield
   {journal} {\bibinfo  {journal} {Physical Review E}\ }\textbf {\bibinfo
  {volume} {70}},\ \bibinfo {pages} {068602} (\bibinfo {year}
  {2004})}\BibitemShut {NoStop}%
\bibitem [{\citenamefont {Shvartsburg}\ \emph {et~al.}(2008)\citenamefont
  {Shvartsburg}, \citenamefont {Marklund}, \citenamefont {Brodin},\ and\
  \citenamefont {Stenflo}}]{PhysRevE.78.016601}%
  \BibitemOpen
  \bibfield  {author} {\bibinfo {author} {\bibfnamefont {A.~B.}\ \bibnamefont
  {Shvartsburg}}, \bibinfo {author} {\bibfnamefont {M.}~\bibnamefont
  {Marklund}}, \bibinfo {author} {\bibfnamefont {G.}~\bibnamefont {Brodin}}, \
  and\ \bibinfo {author} {\bibfnamefont {L.}~\bibnamefont {Stenflo}},\
  }\bibfield  {title} {\enquote {\bibinfo {title} {Superluminal tunneling of
  microwaves in smoothly varying transmission lines},}\ }\href {\doibase
  10.1103/PhysRevE.78.016601} {\bibfield  {journal} {\bibinfo  {journal}
  {Physical Review E}\ }\textbf {\bibinfo {volume} {78}},\ \bibinfo {pages}
  {016601} (\bibinfo {year} {2008})}\BibitemShut {NoStop}%
\bibitem [{\citenamefont {Qiu}\ and\ \citenamefont
  {Zouhdi}(2007)}]{PhysRevB.75.196101}%
  \BibitemOpen
  \bibfield  {author} {\bibinfo {author} {\bibfnamefont {Cheng-Wei}\
  \bibnamefont {Qiu}}\ and\ \bibinfo {author} {\bibfnamefont {Said}\
  \bibnamefont {Zouhdi}},\ }\bibfield  {title} {\enquote {\bibinfo {title}
  {Comment on ``negative refractive index in gyrotropically magnetoelectric
  media''},}\ }\href {\doibase 10.1103/PhysRevB.75.196101} {\bibfield
  {journal} {\bibinfo  {journal} {Phys. Rev. B}\ }\textbf {\bibinfo {volume}
  {75}},\ \bibinfo {pages} {196101} (\bibinfo {year} {2007})}\BibitemShut
  {NoStop}%
\bibitem [{\citenamefont {Pauli}\ and\ \citenamefont
  {Villars}(1949)}]{RevModPhys.21.434}%
  \BibitemOpen
  \bibfield  {author} {\bibinfo {author} {\bibfnamefont {Wolfgang}\
  \bibnamefont {Pauli}}\ and\ \bibinfo {author} {\bibfnamefont
  {F.}~\bibnamefont {Villars}},\ }\bibfield  {title} {\enquote {\bibinfo
  {title} {On the invariant regularization in relativistic quantum theory},}\
  }\href {\doibase 10.1103/RevModPhys.21.434} {\bibfield  {journal} {\bibinfo
  {journal} {Reviews of Modern Physics}\ }\textbf {\bibinfo {volume} {21}},\
  \bibinfo {pages} {434--444} (\bibinfo {year} {1949})}\BibitemShut {NoStop}%
\bibitem [{\citenamefont {Feynman}(1949)}]{PhysRev.76.769}%
  \BibitemOpen
  \bibfield  {author} {\bibinfo {author} {\bibfnamefont {Richard~Phillips}\
  \bibnamefont {Feynman}},\ }\bibfield  {title} {\enquote {\bibinfo {title}
  {Space-time approach to quantum electrodynamics},}\ }\href {\doibase
  10.1103/PhysRev.76.769} {\bibfield  {journal} {\bibinfo  {journal} {Physical
  Review}\ }\textbf {\bibinfo {volume} {76}},\ \bibinfo {pages} {769--789}
  (\bibinfo {year} {1949})}\BibitemShut {NoStop}%
\bibitem [{\citenamefont {Schwinger}(1951)}]{PhysRev.82.664}%
  \BibitemOpen
  \bibfield  {author} {\bibinfo {author} {\bibfnamefont {Julian}\ \bibnamefont
  {Schwinger}},\ }\bibfield  {title} {\enquote {\bibinfo {title} {On gauge
  invariance and vacuum polarization},}\ }\href {\doibase
  10.1103/PhysRev.82.664} {\bibfield  {journal} {\bibinfo  {journal} {Physical
  Review}\ }\textbf {\bibinfo {volume} {82}},\ \bibinfo {pages} {664--679}
  (\bibinfo {year} {1951})}\BibitemShut {NoStop}%
\bibitem [{\citenamefont {Tsai}\ and\ \citenamefont
  {Erber}(1974)}]{PhysRevD.10.492}%
  \BibitemOpen
  \bibfield  {author} {\bibinfo {author} {\bibfnamefont {{Wu-yang}}\
  \bibnamefont {Tsai}}\ and\ \bibinfo {author} {\bibfnamefont {Thomas}\
  \bibnamefont {Erber}},\ }\bibfield  {title} {\enquote {\bibinfo {title}
  {Photon pair creation in intense magnetic fields},}\ }\href {\doibase
  10.1103/PhysRevD.10.492} {\bibfield  {journal} {\bibinfo  {journal} {Physical
  Review D}\ }\textbf {\bibinfo {volume} {10}},\ \bibinfo {pages} {492--499}
  (\bibinfo {year} {1974})}\BibitemShut {NoStop}%
\bibitem [{\citenamefont {Tsai}\ and\ \citenamefont
  {Erber}(1975)}]{PhysRevD.12.1132}%
  \BibitemOpen
  \bibfield  {author} {\bibinfo {author} {\bibfnamefont {{Wu-yang}}\
  \bibnamefont {Tsai}}\ and\ \bibinfo {author} {\bibfnamefont {Thomas}\
  \bibnamefont {Erber}},\ }\bibfield  {title} {\enquote {\bibinfo {title}
  {Propagation of photons in homogeneous magnetic fields: Index of
  refraction},}\ }\href {\doibase 10.1103/PhysRevD.12.1132} {\bibfield
  {journal} {\bibinfo  {journal} {Physical Review D}\ }\textbf {\bibinfo
  {volume} {12}},\ \bibinfo {pages} {1132--1137} (\bibinfo {year}
  {1975})}\BibitemShut {NoStop}%
\bibitem [{\citenamefont {Chiao}(1993)}]{PhysRevA.48.R34}%
  \BibitemOpen
  \bibfield  {author} {\bibinfo {author} {\bibfnamefont {Raymond~Y.}\
  \bibnamefont {Chiao}},\ }\bibfield  {title} {\enquote {\bibinfo {title}
  {Superluminal (but causal) propagation of wave packets in transparent media
  with inverted atomic populations},}\ }\href {\doibase
  10.1103/PhysRevA.48.R34} {\bibfield  {journal} {\bibinfo  {journal} {Phys.
  Rev. A}\ }\textbf {\bibinfo {volume} {48}},\ \bibinfo {pages} {R34--R37}
  (\bibinfo {year} {1993})}\BibitemShut {NoStop}%
\bibitem [{\citenamefont {Diener}(1996)}]{Diener1996327}%
  \BibitemOpen
  \bibfield  {author} {\bibinfo {author} {\bibfnamefont {G.}~\bibnamefont
  {Diener}},\ }\bibfield  {title} {\enquote {\bibinfo {title} {Superluminal
  group velocities and information transfer},}\ }\href {\doibase
  10.1016/S0375-9601(96)00767-0} {\bibfield  {journal} {\bibinfo  {journal}
  {Physics Letters A}\ }\textbf {\bibinfo {volume} {223}},\ \bibinfo {pages}
  {327 -- 331} (\bibinfo {year} {1996})}\BibitemShut {NoStop}%
\bibitem [{\citenamefont {Chiao}\ and\ \citenamefont
  {Milonni}(2002)}]{Chiao:02}%
  \BibitemOpen
  \bibfield  {author} {\bibinfo {author} {\bibfnamefont {Raymond~Y.}\
  \bibnamefont {Chiao}}\ and\ \bibinfo {author} {\bibfnamefont {Peter~W.}\
  \bibnamefont {Milonni}},\ }\bibfield  {title} {\enquote {\bibinfo {title}
  {Fast light, slow light},}\ }\href {\doibase 10.1364/OPN.13.6.000026}
  {\bibfield  {journal} {\bibinfo  {journal} {Optics \& Photonics News}\
  }\textbf {\bibinfo {volume} {13}},\ \bibinfo {pages} {26--30} (\bibinfo
  {year} {2002})}\BibitemShut {NoStop}%
\bibitem [{\citenamefont {Greenberger}\ and\ \citenamefont
  {Svozil}(2005)}]{svozil-greenberger-2005}%
  \BibitemOpen
  \bibfield  {author} {\bibinfo {author} {\bibfnamefont {Daniel~M.}\
  \bibnamefont {Greenberger}}\ and\ \bibinfo {author} {\bibfnamefont {Karl}\
  \bibnamefont {Svozil}},\ }\bibfield  {title} {\enquote {\bibinfo {title}
  {Quantum theory looks at time travel},}\ }in\ \href@noop {} {\emph {\bibinfo
  {booktitle} {Quo Vadis Quantum Mechanics?}}},\ \bibinfo {editor} {edited by\
  \bibinfo {editor} {\bibfnamefont {S.~Dolev}\ \bibnamefont {A.~Elitzur}}\ and\
  \bibinfo {editor} {\bibfnamefont {N.}~\bibnamefont {Kolenda}}}\ (\bibinfo
  {publisher} {Springer},\ \bibinfo {address} {Berlin},\ \bibinfo {year}
  {2005})\ pp.\ \bibinfo {pages} {63--72},\ \Eprint
  {http://arxiv.org/abs/quant-ph/0506027} {quant-ph/0506027} \BibitemShut
  {NoStop}%
\bibitem [{\citenamefont {Svozil}(2000)}]{svozil-relrel}%
  \BibitemOpen
  \bibfield  {author} {\bibinfo {author} {\bibfnamefont {Karl}\ \bibnamefont
  {Svozil}},\ }\bibfield  {title} {\enquote {\bibinfo {title} {Relativizing
  relativity},}\ }\href {\doibase 10.1023/A:1003600519752} {\bibfield
  {journal} {\bibinfo  {journal} {Foundations of Physics}\ }\textbf {\bibinfo
  {volume} {30}},\ \bibinfo {pages} {1001--1016} (\bibinfo {year} {2000})},\
  \Eprint {http://arxiv.org/abs/quant-ph/0001064} {quant-ph/0001064}
  \BibitemShut {NoStop}%
\bibitem [{\citenamefont {Svozil}(2002)}]{svozil-2001-convention}%
  \BibitemOpen
  \bibfield  {author} {\bibinfo {author} {\bibfnamefont {Karl}\ \bibnamefont
  {Svozil}},\ }\bibfield  {title} {\enquote {\bibinfo {title} {Conventions in
  relativity theory and quantum mechanics},}\ }\href {\doibase
  10.1023/A:1015017831247} {\bibfield  {journal} {\bibinfo  {journal}
  {Foundations of Physics}\ }\textbf {\bibinfo {volume} {32}},\ \bibinfo
  {pages} {479--502} (\bibinfo {year} {2002})},\ \Eprint
  {http://arxiv.org/abs/quant-ph/0110054} {quant-ph/0110054} \BibitemShut
  {NoStop}%
\bibitem [{\citenamefont {Peres}(1984)}]{peres-84}%
  \BibitemOpen
  \bibfield  {author} {\bibinfo {author} {\bibfnamefont {Asher}\ \bibnamefont
  {Peres}},\ }\bibfield  {title} {\enquote {\bibinfo {title} {Defining
  length},}\ }\href {\doibase 10.1038/312010b0} {\bibfield  {journal} {\bibinfo
   {journal} {Nature}\ }\textbf {\bibinfo {volume} {312}},\ \bibinfo {pages}
  {10} (\bibinfo {year} {1984})}\BibitemShut {NoStop}%
\end{thebibliography}

%

\end{document}